\documentclass[12pt,aps,nofootinbib,amsfonts,superscriptaddress,showpacs,showkeys]{revtex4}
\usepackage{amsfonts,amssymb,amscd,amsmath}
\usepackage{graphicx}
\DeclareMathOperator{\tr}{tr}
\DeclareMathOperator{\Tr}{Tr}

\newcommand{\slashit}[1]{#1 \kern-.45em\slash}

\newcommand{\slashP}{P \kern-.65em\slash }

\begin{document}
\title{ The Nuclear matter stability in a non-local chiral quark model}
\author{Amir H. Rezaeian}
\email{Rezaeian@tphys.uni-heidelberg.de}
\author{Hans-Juergen Pirner}
\email{pir@tphys.uni-heidelberg.de}
\affiliation{Institute for Theoretical Physics, University of Heidelberg,
Philosophenweg 19, D-69120 Heidelberg, Germany}
\date{\today}
\begin{abstract}
We study the stability of the nuclear matter in a non-local
Nambu-Jona-Lasinio model.  We work out the equation of state in a
relativistic Faddeev approach where we take into account the internal
structure of nucleon. We show that the binding energy saturates when
the nucleon as a composite particle made of quarks is incorporated. After truncation of the two-body channels to the
scalar and axial-vector diquarks, a relativistic Faddeev equation for
nucleon bound states in medium is solved in the covariant
diquark-quark picture. We investigate the nucleon properties in the
nuclear medium such as the role of diquarks within the nucleon and the
in-medium modification of the nucleon mass and size.
\end{abstract}
\pacs{12.39.Ki,12.39.Fe,21.65.+f,24.85.+p} 
\keywords{Non-local Nambu-Jona-Lasinio model, nuclear matter stability, quark-diquark structure of the nucleon, quark matter} 
\maketitle
\date{\today}

%\maketitle

\section{Introduction}

Chiral type models such as the NJL model have been successful in
explaining the low energy physics of mesons and nucleons in vacuum
utilizing the concept of the spontaneous chiral symmetry breaking
\cite{njl}. Application of the model to low density, however, has
posed serious problems like the absence of saturation
\cite{sa-1,sa-1-1,sa-2,under2}. This is in contrast to non-chiral models like the Walecka
model \cite{sa-3} which has been phenomenologically successful in describing the
nuclear matter ground state in the mean field approximation.

In chiral models with a Mexican hat potential, the nuclear
medium moves the minimum of the vacuum effective
potential towards smaller values. This implies a smaller curvature, i.e. the scalar
sigma mass decreases. Consequently, the attraction between nucleons
due to sigma meson exchange increases. This effect may destroy the stability
of nuclear matter. On the contrary, in non-chiral models
e.g. the Walecka type model, where the potential is simpler, a stable nuclear matter can be
found.

At the nuclear matter saturation point the Fermi momentum $k_{f}$ and
the pion mass $m_{\pi}$ are of comparable scale ($k_{f}\simeq 262
MeV\simeq 2m_{\pi}$). Therefore, chiral dynamics may play an important
role in a nuclear matter saturation mechanism. Chiral symmetry is an
approximate symmetry for quarks which is not necessarily translated
into a symmetry for nucleon degrees of freedom. Furthermore, the
delta-nucleon mass splitting $\Delta =293$ MeV is also comparable to
the Fermi momentum $k_{f}$ at nuclear matter saturation. It has been
shown that the main mechanism behind the delta-nucleon mass splitting
stems from the fact that the delta and the nucleon have different
internal quark structure\footnote{The Delta is made of only axial
vector diquarks while nucleons can be constructed from both the scalar
and axial vector diquarks.} \cite{delta1,njln3}. Therefore one may
naturally be led to ask if the internal structure of nucleon is
important in order to describe the empirical nuclear matter ground
state. Quark degrees of freedom have also been important in order to
describe deep-inelastic scattering at momentum transfers of several
GeV and the EMC effect \cite{p1,sa-4,fad-new}.

Moreover, at the moment a direct calculation of the many nucleons
system based on QCD itself is not feasible. Therefore, effective quark
theories are the best tools to develop a microscopic understanding of
a single nucleon together with nuclear and quark matter and their phase
structure in a unified framework.

It has been shown that a non-local covariant extension of the
NJL model inspired by the instanton liquid model
\cite{non1} can lead to quark confinement for acceptable values of
the parameters \cite{pb}.  There are other advantages of the non-local
version of the model over the local NJL model, for example: the
constituent quark mass is momentum-dependent, as also found in lattice
calculation \cite{lat-m}. The regulator makes the theory finite to all
orders in the loop expansion and leads to small next-to-leading order
corrections \cite{pb-2}. This model has been phenomenologically very
successful to describe mesons \cite{pb,pb-2,a1} and baryons
\cite{a3,me1} in vacuum. More applications can be found in Refs.~\cite{a2,non-s2}. Here, we employ this model to investigate the role of
the internal dynamics of nucleon in a nuclear matter environment. We
explicitly construct nucleon degrees of freedom by binding diquarks
and quarks with relativistic Faddeev equations to describe nuclear
matter. The relativistic Faddeev approach to describe the nucleon as a
composite object has been very successful
\cite{under2,njln1,fad-new,njln3,o1,o2,me1,me2}.

Very recently Bentz and Thomas \cite{under2} have shown that local NJL
models may lead to saturating nuclear matter when the quark confinement
effect is incorporated to avoid unphysical thresholds.  Although the
idea is interesting, there may be some short-coming in their
approach. Quark confinement in their prescription is incorporated by
introducing an infrared cutoff in the model. This way quark
confinement is permanent and there is no confinement-deconfinement 
transition in their model. In their hybrid approach,
the nucleon is described as a relativistic bound state in the static approximation\footnote{In the static approximation of the Faddeev equation the quark exchange kernel is taken momentum independent, which  may give rise 
to an unphysical in-medium singularity due to a reduction of the
constituent quark mass.}  of a scalar diquark and quark. It has been
already noticed that the attractive axial-vector diquark plays an
important role in description of a single nucleon
\cite{njln3,me1}. Therefore, in order to treat the internal structure of the nucleon and its modification 
in the medium adequately, we will take into account both the scalar and the
axial-vector diquark channels and solve the Faddeev equation without
invoking the static approximation.

We start by investigating the role of diquarks in the bound
nucleon. The in-medium modification of the nucleon mass and the
nucleon size will be calculated. We obtain the nuclear matter equation
of state when nucleon is taken as a composite object made of diquark
and quark.

This paper is organised as follows: In Sec.~II we introduce the model,
discuss its pionic sector and fix the parameters. In Sec.~III we set up
the description of the nucleon based on a diquark-quark picture in the
relativistic Faddeev approach. In Sec.~IV we consider nuclear
matter with composite nucleons and present our numerical
results. Finally, a summary and outlook is given in Sec.~VI.

\section{Formalism and solution in the vacuum and quark matter\label{sec:model}}
We consider a non-local NJL model Lagrangian in terms of quark degrees of freedom with $SU(2)_{f}\times SU(3)_{c}$ symmetry
\begin{equation}
\mathcal{L}=\bar{\psi}(i\slashit\partial-m_{c})\psi+\frac{1}{2}g_{\pi} j_{\alpha}(x)j_{\alpha}(x),\label{e1}
\end{equation}
where $m_{c}$ is the current quark mass of the $u$ and $d$ quarks. First, we
restrict the interaction terms to four-quark interaction vertices in
the quark-antiquark channels. We will add in section III the diquark
channels in order to model the quark sub-structure of the nucleon in
the medium. The current $j_{\alpha}(x)$ has the scalar
($0^{+},T=0$) and pseudoscalar ($0^{-},T=1$) components 
\begin{equation}\label{n1}
j_{\alpha}(x)=\int
d^{4}x_{1}d^{4}x_{3}F(x-x_{3})F(x_{1}-x)\overline{\psi}(x_{1})\Gamma_{\alpha}\psi(x_{3}),
\end{equation}
with $\Gamma_{\alpha}=(1,i\gamma_{5}\vec{\tau})$. The ${\tau_{i}}$'s are
flavour $SU(2)$ matrices with
$\tr(\tau_{i}\tau_{j})=2\delta_{ij}$. Notice that there are different
approaches \cite{asy,non-bir} to introduce non-locality into the
interactions. We use the four-way separability of the non-local
interaction which is also present in the instanton liquid model
\cite{sep-i}. This assumption considerably simplifies the calculation. Non-locality
also emerges naturally in the presence of other gluonic field configurations
within the QCD vacuum \cite{gluon} and in models based on Schwinger-Dyson
resummation techniques \cite{asy}.

In the medium the repulsive vector mesons are important. For
simplicity we add a local vector field interaction
$-g_{v}(\overline{\psi}\gamma^{\mu}\psi)^{2}$ to the
Lagrangian. Locality of this term is permissible as long as the
interactions in Lagrangian are not fixed by some underlying theory
through a Fierz transformation. Upon standard bosonization we obtain
\begin{eqnarray}
\mathcal{L}&=&\int d^{4}x_{1}d^{4}x_{3}\overline{\psi}(x_{1})\Big[\left(i\slashit\partial-m_{c}\right)\delta(x_{1}-x_{3})\delta(x_{3}-x)+F(x-x_{3})F(x_{1}-x)\left(\sigma(x)+\vec{\pi(x)}i\gamma_{5}\vec{\tau} \right)\nonumber\\
&-&\gamma^{\mu}\omega_{\mu}(x)\delta(x_{1}-x_{3})\delta(x_{3}-x)\Big]\psi(x_{3})-\frac{1}{2g_{\pi}}[\sigma^{2}(x)+\pi^{2}(x)]+\frac{1}{4g_{v}}\omega_{\mu}(x)\omega^{\mu}(x).\label{e2}\
\end{eqnarray}
Lagrangians (\ref{e1}) and (\ref{e2}) are equivalent in the classical limit. The equations of motion for bosonic fields lead to   
\begin{eqnarray}
\left(\sigma(x), \vec{\pi}(x)\right)&=&g_{\pi}\int d^{4}x_{1}d^{4}x_{3}F(x-x_{3})
F(x_{1}-x)\overline{\psi}(x_{1})\left(1,i\gamma_{5}\vec{\tau}\right)\psi(x_{3}),\nonumber\\
\omega^{\mu}(x)&=&2g_{v}\overline{\psi}(x)\gamma^{\mu}\psi(x).\label{e3}\
\end{eqnarray} 
The Lagrangian (\ref{e1}) can be immediately recovered from the bosonized version
Eq.~(\ref{e2}) by substituting Eq.~(\ref{e3}) into Eq.~(\ref{e2}).

We define the Fourier transform of the form factor by
\begin{equation}
F(x-x_{i})=\int \frac{d^{4}p}{(2\pi)^{4}} e^{-i(x-x_{i})\cdot p}f(p). \label{forfa}
\end{equation}
The function $f(p)$ influences the momentum distribution of quarks in
the non-perturbative vacuum and will be defined later. Next, we
perform a mean field approximation by expanding the meson fields
around their expectation values $\overline{\sigma}=\langle
\sigma\rangle$, $\overline{\omega}^{0}=\langle\omega^{0}\rangle$ neglecting meson fluctuations. The
mean values of the pion field and space components of the vector field
vanish in vacuum and for a baryonic matter at rest due to
symmetry. The mean field Lagrangian in momentum space can then be
written as
\begin{equation}
\mathcal{L}=\overline{\psi}(p)\left(i\slashit p- M(p)-\gamma^{0}\overline{\omega}_{0}\right)\psi(p)-\frac{\overline{\sigma}^2}{2g_{\pi}}+\frac{\overline{\omega}_{0}^{2}}{4g_{v}}, \label{mean-l}
\end{equation}
where the momentum-dependent quark mass and quark propagator are given by 
\begin{eqnarray}
M(p)=m_{c}+f^{2}(p)\overline{\sigma}, \label{mass}\\
S^{-1}(p)=\slashit p-M(p)-\gamma^{0}\overline{\omega}_{0}. \label{pro}\
\end{eqnarray}
We treat the quark fields at one-loop level and the meson fields at
tree level. This approximation is consistent with the leading-order
behaviour of the model in a $1/N_{c}$ expansion. Within this
approximation the effective potential can be calculated:
\begin{equation}
\Omega= i\gamma_{q}\int \frac{d^{4}k}{(2\pi)^4}\ln \left(k^{2}-M^{2}(k)\right)+\frac{\overline{\sigma}^2}{2g_{\pi}}-\frac{\overline{\omega}_{0}^{2}}{4g_{v}},   \label{vac1-0}
\end{equation}
where $\gamma_{q}=2N_{c}N_{f}$ is the degeneracy factor for quarks. A
gap equation is obtained by requiring
$\frac{\partial\Omega}{\partial\overline{\sigma}}=0$:
\begin{equation}
\overline{\sigma}=2g_{\pi}\gamma_{q}i\int \frac{d^{4}k}{(2\pi)^4}\frac{M(k)f^{2}(k)}{k^{2}-M^{2}(k)}, \label{sel1-0}
\end{equation}
where the dynamical quark mass is given in Eq.~(\ref{mass}).

Our model contains four parameters: the current quark mass $m_{c}$,
the cutoff ($\Lambda$) hidden in the form factor $f(p)$ and the coupling constants $g_{\pi}$ ,
$g_{v}$. The vector field coupling $g_{v}$ is treated as a free
parameter and will be adjusted in the medium. We first fix $g_{\pi}$ and
$m_{c}$ for arbitrary values of $\Lambda $ by fitting the pion mass
$m_{\pi}$ and the pion decay constant $f_{\pi}$ to their empirical values at zero
baryon density. In this way, we can consider the entire parameter space of
the model.  The pion mass is given by the zero of the 1PI two-point
function $\Pi_{\pi\pi}(q)$ for the pion which in random phase
approximation has the following form
\begin{equation}
\Pi_{\pi\pi}(q)=\frac{1}{g_{\pi}}+ i \Tr \int
 \frac{d^{4}k}{(2\pi)^{4}}f^{2}(k)\gamma_{5}\tau_{i}S(k)\gamma_{5}\tau_{i}S(q+k)f^{2}(q+k)=0, \label{lo1}
\end{equation}
where $q$ denotes the total momentum of the quark-antiquark
pair\footnote{The symbol $\Tr$ denotes a trace over flavour, colour
and Dirac indices.}. The pion decay constant $f_{\pi}$ is obtained from the coupling of the
pion to the axial-vector current \cite{pb}
\begin{eqnarray}
f_{\pi}&=&\frac{ig_{\pi \bar{q}q}}{m_{\pi}^{2}}\int
\frac{d^{4}p}{(2\pi)^{4}}\Tr[\slashit{q}\gamma_{5}\frac{\tau_{a}}{2}(S(p_{-}))\gamma_{5}\tau_{a}(S(p_{+}))]f(p_{-})f(p_{+})\nonumber\\
&-&\frac{g_{\pi}g_{\pi \bar{q}q}}{2m^{2}_{\pi}}\int
\frac{d^{4}k}{(2\pi)^{4}}\Tr[S(k)]\int
\frac{d^{4}p}{(2\pi)^{4}}\Tr[\tau^{a}\gamma_{5}f(p_{-})f(p_{+})S(p_{-})\gamma_{5}\tau_{a}S(p_{+})]\nonumber\\
&&\times
[f^{2}(k)\left(f^{2}(p_{+})+f^{2}(p_{-})\right)-f(p_{+})f(p_{-})f(k)\left(f(k+q)+f(k-q)\right)], \label{fpi}\
\end{eqnarray}
where $g_{\pi \bar{q}q}$ is the pion-quark-antiquark coupling constant
and related to the corresponding loop integral Eq.~(\ref{lo1}) by 
$g_{\pi\bar{q}q}^{-2}=(\frac{d\Pi_{\pi\pi}}{dq^{2}})|_{q^{2}=m^{2}_{\pi}}$. We define $p_{\pm}=p\pm\frac{1}{2}q$.  Notice that due to
non-locality the one-pion-to-vacuum matrix element gets an additional
contribution. This extra term is very important to maintain the
Gell-Mann-Oakes-Renner relation \cite{pb} and makes a significant
numerical contribution.

In our numerical treatment of the model, we evaluate loop-integrals
like that in Eqs.~(\ref{sel1-0},\ref{lo1},\ref{fpi}) in Euclidean
space\footnote{Notice that here (e.g in Eq.~(\ref{sel1-0})) a standard
transcription rules from Minkowski to Euclidean momentum ($k^{0}\to
ik_{4},\vec k \to-\vec k^{E}$) and reverse has been assumed.
%be easily justified due to the complex pole in dressed quark
%propagator. At the moment there is no other alternative
%available for such a highly non-trivial problem. 
%Here we assume that such a transcription rules can be applied. 
So far this assumption has been phenomenologically correct
\cite{non1,pb,pb-2,a1,me1,non-s2,a2,a3}.}. We choose for the form factor in Eq.~(\ref{forfa}) a Gaussian
function in Euclidean space, 
\begin{equation}
f(p_{E})=\exp(-p_{E}^{2}/\Lambda^{2}), \label{f-f}
\end{equation}
where $\Lambda$ is the cutoff of the theory. This choice respects
Poincar\'e invariance and for certain values of the parameters it
leads to the quark ''confinement'' in the sense that the dressed quark
propagator has no poles at positive $p^{2}$ in Minkowski space and
consequently, quarks do not appear as asymptotic states. Quark
''confinement'' can occur if the following relation is satisfied between
the parameters of the model
\begin{equation} \label{n4}
\frac{\overline{\sigma}}{\sqrt{m^{2}_{c}+\Lambda^{2}}-m_{c}} >
\frac{1}{2}\exp\left(-\frac{(\sqrt{m^{2}_{c}+\Lambda^{2}}+m_{c})^{2}}{2\Lambda^{2}}\right).
\end{equation}
Notice that the quark propagator has many pairs of complex poles, both
for confining and non-confining parameter sets which can be conceived
as a remnant of the underlying confinement. Confinement in this model
has direct consequence for the pion, but condition Eq.~(\ref{n4}) is
not derived by using any assumption about pions.
\begin{figure}[!tp]
       \centerline{\includegraphics[width=14 cm] {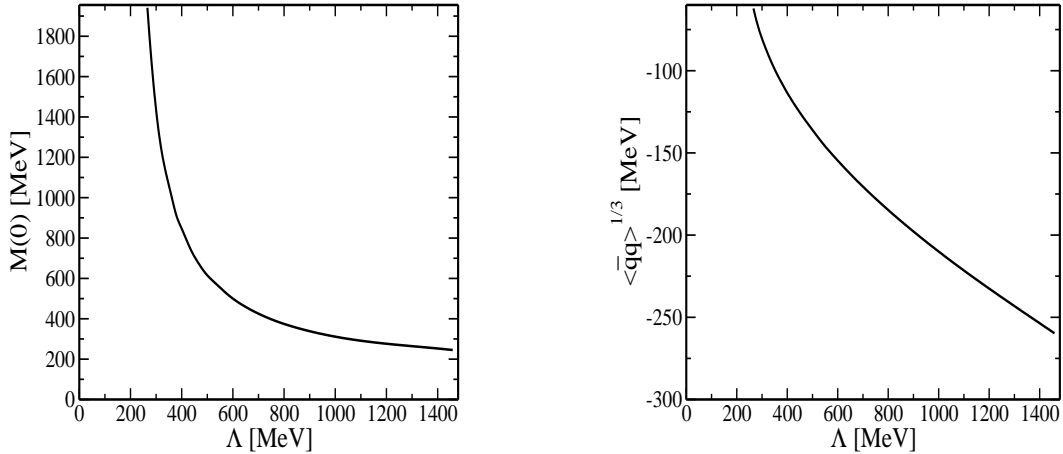}}
       \caption{The zero-momentum constituent quark (left panel) and
       quark condensate (right panel) as a function of cutoff for
       fixed $f_{\pi}= 93$ MeV and $m_{\pi}=140$ MeV when $m_{c}\neq
       0$. }
\end{figure}
\begin{figure}[!tp]
       \centerline{\includegraphics[width=7 cm] {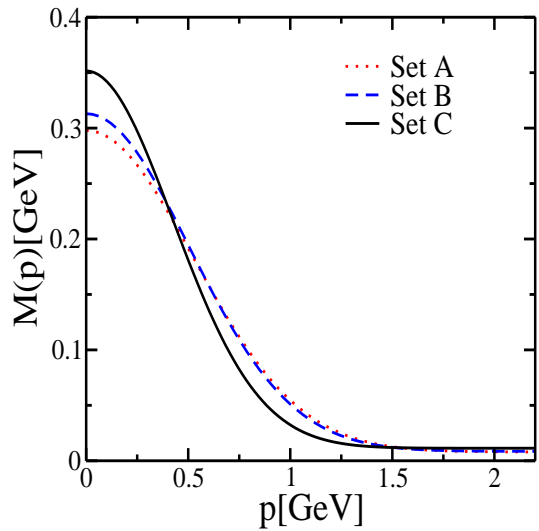}}
       \caption{The constituent quark mass $M(p)$ as a function of momentum $p$ for the three parameter sets given in Table 1. } 
\end{figure}

\begin{table}
%\centering
%\begin{center}
\caption{The parameters $\Lambda$, $g_{\pi}$ and $m_{c}$ for the sets $A$, $B$ and $C$, 
fitted to $f_{\pi}=92.4$ MeV and $m_{\pi}=139.6$ MeV. The resulting
values of the constituent quark mass $M(0)$ (in brackets we give the values for zero current
quark mass) and the corresponding quark condensate for the different
parameter sets are also given.\label{tab:modpar}}
\begin{ruledtabular}
\begin{tabular}{llll}
Parameter &set A & set B & set C  \\
\hline
$\Lambda$ (MeV) & 1046.8 &1008.1 &847.8 \\
$g_{\pi} (\text{GeV}^{-2})$ &31.6 &35.2 &55.80 \\
$m_{c}$ (MeV) & 7.9(0) &8.5(0) &11.13(0) \\
$M(0)$ (MeV) & 297.9(250)& 313(264.5) &351.6(300)  \\
$-\langle \bar{q}q\rangle^{1/3}$ (MeV) &215(208) & 212(205) & 191(186) \\
\end{tabular}
\end{ruledtabular}
%\end{table}
%\begin{table}
\caption{The first two sets of  poles of the quark propagator $\pm M_{R}\pm iM_{I}$ in
Minkowski space with $m_{c}\neq 0$ . \label{tab:quarkpoles}}
\centering
\begin{ruledtabular}
\begin{tabular}{lll}
set A  & set B & set C \\
\hline
$\pm 391$ MeV&$\pm 507 \pm i55$ MeV &$\pm 408 \pm i238$ MeV\\
$\pm 675$ MeV&$\pm 1530 \pm i253$ MeV &$\pm 1575\pm i307$ MeV\\
\end{tabular}
\end{ruledtabular}
\end{table}

The emergence of complex poles for quark propagators was also noticed
in Schwinger-Dyson equation studies in QED and QCD
\cite{sde-p,sde-new,sde-qe}. It has been recently shown that baryons
become more compact in the presence of imaginary poles in the quark
propagator \cite{me1}. In another study it has been noticed that the
mass-like singularities are located on the time axis if one removes
the confining potential in QED in 2+1D, and the mass-like
singularities move from the time axis to complex momenta if there is a
confining potential \cite{sde-qe}.

We refer to appendix for details of numerical evaluation of loop
integrals as of Eq.~(\ref{lo1}).  Solutions of the gap equation
(\ref{sel1-0}) are shown in Fig.~1. In the left panel the constituent
mass at zero momentum $M(0)$ is shown as a function of the cutoff. It
is obvious that for very small cutoff, there is no solution for the
gap equation. On the right panel of Fig.~1, we show the corresponding
values of the quark condensate.  These values are within the limits
extracted from QCD sum rules
\cite{qcdsum} and also lattice calculation \cite{lat-con}. Notice that the trend of the model parameters in
Fig.~1 is slightly different to that obtained in the local NJL model
with a sharp cutoff \cite{bub1}. In contrast to the local NJL model,
the dynamical quark mass Eq.~(\ref{mass}) is momentum dependent (see
Fig.~2) and follows a very similar trend to lattice simulations
\cite{lat-m}. Fig.~2 shows that at low virtualities the quark mass is
close to the constituent mass while at large virtualities it
approaches the current mass.

We analyse three sets of parameters, as indicated in Table
\ref{tab:modpar}. Set $A$ is called the non-confining parameter set since it does not contain imaginary poles of the quark propagator.
Sets $B$ and $C$ lead to the quark ''confinement'' in our convention
with complex poles. We define the pole mass to be given through the
equation 
\begin{equation}
P_{pole}^{2}=M^{2}(P_{pole}),\label{on-shell}
\end{equation}
where pole solutions are denoted $P_{pole}=\pm M_{R}\pm iM_{I}$.
Equation (\ref{on-shell}) is a non-linear equation and can be solved
numerically.  The position of the quark poles are given in Table
\ref{tab:quarkpoles} where it is seen that for the confining sets the quark poles lie in the complex
plane. In contrast to the local NJL model, a non-zero current quark
mass of order $8.5$ MeV leads to an increase of about $48.5$ MeV in the
zero-momentum dynamical quark mass (see Table 1 set B). It is also
noted that the real part of the pole mass is bigger than the
zero-momentum constituent quark mass $M(0)$ for both the confining and
the non-confining sets, see Tables 1,2.

\section{Nucleon internal structure}
 In order to have a physically realistic description of matter at low
 density, one has to construct nucleons and nucleon matter from the
 quark degrees of freedom since quarks are confined at very low
 density.  In this section, we take into account the quark structure of the
nucleon in a diquark-quark picture. We first build up a single nucleon
out of the quark degrees of freedom by solving the relativistic
Faddeev equation, then (in the next section) we construct nuclear
matter in a mean field approximation by means of these individual
nucleons. In this way, any reference to the quarks will be naturally
hidden in the nucleons.

In the vacuum baryons and diquarks have already been solved in this model by one of the authors \cite{me1}. Here, we shortly recapitulate
the procedure to construct the nucleon as a diquark-quark bound state, more
details can be found in \cite{me1,me2}.

In order to describe the nucleon as a bound state of a diquark and a quark, we
firstly introduce diquarks in the model.  We truncate
the quark-quark interaction to the scalar $(0^{+},T=0, \overline{3})$
and the axial vector $(1^{+}, T=1,\overline{3})$ colour antitriplet,

\begin{equation}
\mathcal{L}_{Is,a}=g_{s}\overline{J}_{s}(x)J_{s}(x)+g_{a}\overline{J}_{a}(x)J_{a}(x),
\end{equation}
where the currents are defined
\begin{eqnarray}\label{n1-rp}
\overline{J}_{s}(x)&=&\int
d^{4}x_{1}d^{4}x_{3}f(x-x_{3})f(x_{1}-x)\overline{\psi}(x_{1})\big[\gamma_{5}C\tau_{2}\beta^{A}\big]\overline{\psi}^{T}(x_{3}),\nonumber\\
J_{s}(x)&=&\int
d^{4}x_{2}d^{4}x_{4}f(x-x_{4})f(x_{2}-x)\psi^{T}(x_{2})\big[C^{-1}\gamma_{5}\tau_{2}\beta^{A}\big]\psi(x_{4}),\nonumber\\
\overline{J}_{a}(x)&=&\int
d^{4}x_{1}d^{4}x_{3}f(x-x_{3})f(x_{1}-x)\overline{\psi}(x_{1})\big[\gamma_{\mu}C\tau_{i}\tau_{2}\beta^{A}\big]\overline{\psi}^{T}(x_{3}),\nonumber\\
J_{a}(x)&=&\int
d^{4}x_{2}d^{4}x_{4}f(x-x_{4})f(x_{2}-x)\psi^{T}(x_{2})\big[C^{-1}\gamma^{\mu}\tau_{2}\tau_{i}\beta^{A}\big]\psi(x_{4}).\
\end{eqnarray}
The object $C=i\gamma_{2}\gamma_{5}$ is the charge conjugation matrix.
The matrices $\beta^{A}=\sqrt{3/2} \lambda^{A}(A=2, 5, 7)$ project
onto the colour $\overline{3}$ channel with normalisation
$\tr(\beta^{A}\beta^{A'})=3\delta^{AA'}$. The couplings 
$g_{s}$ and $g_{a}$ specify the strength in the scalar and
axial-vector diquark channels, respectively. We fix them by the
empirical nucleon mass. The form factor in the diquark sector is
assumed to be the same as in the meson sector.

\begin{figure}[!tp]
       \centerline{\includegraphics[width=16 cm] {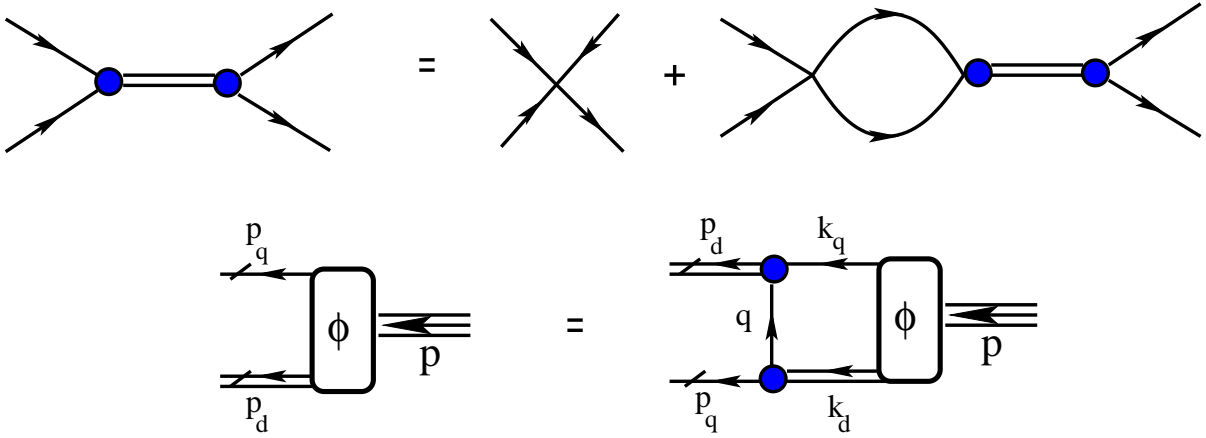}}
       \caption{Top panel: A graphical representation of the
       Bethe-Salpeter equation for the $qq$ $T$-matrix in RPA
       approximation. The solid lines denote the dressed quark
       propagators Eq.~(\ref{pro}) and double lines denote diquark
       propagators $D^{5},D^{\mu\nu}$. The filled circles denote the
       diquark vertices $\chi^{5},\chi^{\mu\nu}$. Down panel: The
       coupled effective Bethe-Salpeter equation for the nucleon
       vertex function $\phi$ is defined by amputating the external quark
       and diquark propagator from the wave function
       $\phi=S^{-1}D^{-1}\psi$.}
\end{figure}

In the ladder approximation the
quark-quark $T$-matrix near the pole can be parametrised as \cite{me1,o2,me2}
\begin{equation}
T(p_{1},p_{2};k_{1},k_{2})=T(p,k;q)\simeq \chi^{5}(p,q)D^{5}(q)\bar{\chi}^{5}(k,q)+\chi^{\mu}(p,q)D^{\mu\nu}(q)\bar{\chi}^{\nu}(k,q),
\label{new-ve}
\end{equation}
where $q=p_{1}+p_{2}=k_{1}+k_{2}$ is the total momentum of the quark-quark pair and $p,k$ are the relative momenta within the diquarks.    
The momentum-dependent scalar $\chi^{5}$, the axial $\chi^{\mu}$ diquark vertex
functions, their adjoint functions $\bar{\chi}^{5}$ and $\bar{\chi}^{\mu}$
and the corresponding propagators $D^{5}(q)$ and
$D^{\mu\nu}(q)$ can be obtained from solving the Bethe-Salpeter
equation \cite{me1} for $qq$ scattering matrix shown in Fig.~3. 
%\begin{eqnarray}
%\chi^{5}(p,q)&=&g_{dsqq}(\gamma^{5}C)\sqrt{2 V^{s}(q)}f(p+(1-\sigma)q)f(-p+\sigma q),\nonumber\\
%\chi^{\mu}(p,q)&=&g_{daqq}(\gamma^{\mu}C)\sqrt{2 V^{a}(q)}f(p+(1-\sigma)q)f(-p+\sigma q).\nonumber\\
%D^{5}(q) &=&  \frac{1}{q^{2}-M^{2}_{ds}}, \hspace{2cm} D^{\mu\nu}(q)=\frac{g^{\mu\nu}-q^{\mu}q^{\nu}/q^{2}}{q^{2}-M^{2}_{da}},\label{di1} \
%\end{eqnarray}
%where $V^{s,a}(q) is the ratio between the exact $T$-matrix and
%one-pole approximation and include the off-shell correction of the
%diquark $T$-matrix around the diquark pole solution (this improves the
%one-shell approximation

The scalar and the axial diquark masses $M_{ds}$ and $M_{da}$ are obtained
as the position of the pole of the corresponding $T$-matrix in the scalar and
axial-vector diquark channels, respectively.

%the 1PI two-point functions\footnote{The bound axial diquark solution
%correspond to a pole of the transverse $T$-matrix. The longitudinal
%one does produce a pole and makes very small contribution compared to
%the transverse part. Moreover, the numerical calculation shows that
%this pole approximation provides enough strength to form a bound
%nucleon.}  ($\Pi_{dsds}(q),\Pi_{dada}^{T}(q)$),

%\begin{eqnarray}
%\Pi_{dsds}(q)&=&1/g_{s}+6i\int\frac{d^{4}k}{(2\pi)^{4}}\tr_{D}[\gamma_{5}S(k)\gamma_{5}S(k+q)]f^{2}(k)f^{2}(q+k)=0,\nonumber\\
%\Pi_{dada}^{T}(q)&=&1/g_{a}+6i (g^{\mu\nu}-q^{\mu}q^{\nu}/q^{2})^{-1}\int\frac{d^{4}k}{(2\pi)^{4}}\tr_{D}[\gamma^{\mu}S(k)\gamma^{\nu}S(k+q)]f^{2}(k)f^{2}(q+k)=0,\label{di-so}\nonumber\\
%\end{eqnarray}
%where the quark propagator is defined in Eq.(\ref{pro}).
%The diquark-quark-quark couplings $g_{dsqq}$ ($g_{dsqq}$) in Eq.~(\ref{di1}) are related to the 1PI two-point function (or to the polarisation function) by
%\begin{equation}
%g^{-2}_{dsqq}=\frac{\Pi_{dsds}(q)}{dq^{2}}|_{q^{2}=M^{2}_{ds}}, \hspace{2cm}
%g^{-2}_{daqq}=\frac{\Pi_{dada}^{T}(q)}{dq^{2}}|_{q^{2}=M^{2}_{da}}.
%\end{equation}
With the diquark, the baryon can be constructed as a bound state of
diquark-quark. We ignore the three-body irreducible graphs. Due to the
separability of the two-body interaction in momentum-space, the
relativistic Faddeev equation can be recast into an effective two-body
Bethe-Salpeter type equation (pictorially shown also in Fig.~3)
\begin{equation}
\int \frac{d^{4}k}{(2\pi)^{4}} G^{-1}(p,k;P) \left(\begin{array}{c}\psi^5 (k,P) \\ \psi^\mu(k,P) \end{array}\right) \label{fad}=0,
\end{equation}
where the $G^{-1}(p,k;K)$ denotes the inverse of the full
quark-diquark $4$-point function and contains the sum of the
disconnected part and the one-quark exchange interaction kernel. After projecting the 
effective diquark-quark Bethe-Salpeter equation in the colour singlet and isospin
$1/2$ channel, one finds
\begin{eqnarray}
G^{-1}&=&(2\pi)^{4}\delta(p-k)S^{-1}(p_{q})\left(\begin{array}{cc}(D^{5}(p_{d}))^{-1}&0\\0&(D^{\nu\mu}(p_{d}))^{-1}\end{array}\right)\nonumber\\
&+&3\left(\begin{array}{cc}\chi^{5}(p_{1},k_{d})S^{T}(q)\bar{\chi}^{5}(p_{2},p_{d})&-\sqrt{3}\chi^{\mu}(p_{1},k_{d})S^{T}(q)\bar{\chi}^{5}(p_{2},p_{d})\\-\sqrt{3}\chi^{5}(p_{1},k_{d})S^{T}(q)\bar{\chi}^{\nu}(p_{2},p_{d})&
-\chi^{\mu}(p_{1},k_{d})S^{T}(q)\bar{\chi}^{\nu}(p_{2},p_{d})\end{array}\right),\label{n9}\
\end{eqnarray}
where $\chi^{5}$ and $\chi^{\mu\nu}$ stand for the Dirac structures of
the diquark vertices introduced in Eq.~(\ref{new-ve}). The dressed quark propagator $S^{-1}(p)$ is
defined in Eq.~(\ref{pro}). The numerical coefficient in kernel of the Faddeev equation
(\ref{n9}) comes from projecting the kernel to the physical baryon
states (colour singlet and isospin half). We define the spectator
quark momentum $p_{q}=\eta P+p$ and the diquark momentum
$p_{d}=(1-\eta)P-p$, where $P$ is the total momenta in the diquark-quark
pair and $\eta\in[0,1]$ is the Mandelstam parameter which parametrises
the relative momenta within the quark-diquark system\footnote{Notice
that observables do not depend on the Mandelstam parameters
$\eta$ \cite{me1,o1,o2,me2}.}. The relative momentum of quarks in the diquarks vertices are
defined as $ p_{1}=p+k/2-(1-3\eta)P/2$ and
$p_{2}=-k-p/2+(1-3\eta)P/2$, and the momentum of the exchanged quark
is $q=-p-k+(1-2\eta)P$. We refer to Appendix for details of numerical methods involved in solving the Faddeev equation (\ref{n9}). 

%\begin{figure}[!tp]
%       \centerline{\includegraphics[width=12 cm] {n.eps}} \caption{The
%       coupled effective Bethe-Salpeter equation for the vertex function $\phi$ defined
%       by amputing the external quark and diquark propagators from the wave function $\phi=S^{-1}D^{-1}\psi$.}
%\end{figure}
The parameters of the models $g_{\pi}, m_{c}$ and $\Lambda$ are fixed by
meson properties i.e. the pion mass and decay in vacuum as described
in section II. We select parameter set C given in Table 1. Now the
only unknown parameters are diquark couplings $g_{s}$ and $g_{a}$ and
vector meson coupling $g_{v}$.  The scalar and axial-vector diquark
couplings are taken as independent parameters. One can find a line in
parameter space of diquark couplings in which a reasonable description
of nucleon does exist, (see Fig.~11 in Ref.~\cite{me1}). In other
words, the interaction is shared between the scalar and the
axial-vector diquark and for small scalar diquark coupling $g_{s}$ one
needs a dominant axial-vector diquark $g_{a}>g_{s}$ and inverse.  We
fix these diquark couplings in vacuum to obtain a nucleon mass of
$940$ MeV. We choose $g_{s}=0.14 g_{\pi}$ and $g_{a}=0.23 g_{\pi}$
(the value of $g_{\pi}$ for set $C$ is given in Table 1) and
corresponds to a scalar diquark with a mass $M_{ds}=705$ MeV and an
axial diquark with a mass $M_{da}=660 $ MeV in vacuum
\cite{me1}.  Note also that it has been shown that the axial-vector
channel is much more important for the confining than non-confining
parameter sets of our model \cite{me1}.  All the results in
the next section are given by the above-mentioned parameter set. The
vector coupling $g_{v}$ is the only free parameter which is left for
the finite density calculation.

\section{Nuclear matter}
In general, using a functional integration technique (similar to
bosonization scheme presented in section II) one can recast the
Lagrangian in terms of hadron degrees of freedom and some auxiliary
fields such as diquarks. The quark and the auxiliary fields
can be integrated out and a chemical potential for the nucleons
should be introduced. Having done that one can then directly apply conventional 
many-body techniques to the effective Lagrangian. Following Bentz and Thomas, we construct baryons out of the
diquark-quark loop by solving the Faddeev equations and solve baryonic matter 
in a mean-field approach\footnote{We use a simple approximation that the nuclear
 matter expectation value of any operator, consists of its expectation
 value in the vacuum and an average over the nucleon
 Fermi-sea of correlated valence nucleon \cite{under2}.}. Since we already have ignored meson
loops, we do not keep diquark loops as well\footnote{Diquarks are alike mesons boson and furthermore, our
consideration is limited to a low-density region where diquarks do not
condensate.}. The vacuum contribution of quark fields is kept at
one-loop level. Therefore, the effective potential $\Omega(\rho)$ regularised by subtracting its corresponding value at zero density
can be written as
\begin{equation}
\Omega(\rho)=\Omega_{\text{Vac}}+\Omega_{\text{Med}},
\end{equation}
the contribution due to vacuum part $\Omega_{\text{Vac}}$ is  
\begin{equation}
\Omega_{\text{Vac}}=i\gamma_{q}\int \frac{d^{4}k}{(2\pi)^4}\ln\frac{k^{2}-M^{2}(k)}{k^{2}-M_{0}^{2}(k)}+\frac{\overline{\sigma}^2-\overline{\sigma}_{0}^2}{2g_{\pi}}-\frac{\overline{\omega}_{0}^{2}}{4g_{v}}. \label{vac1}
\end{equation}
The medium part contains the nucleons and is defined
\begin{equation}
\Omega_{\text{Med}}=\gamma_{N}\int \frac{d^{3}k}{(2\pi)^3} E_{N}(k)n_{N}(k), \label{med2}
\end{equation}
where $\gamma_{N}=4$ is the spin-isospin degeneracy factor for the
nucleon. The energy spectrum of a single nucleon has a simple form
$E_{N}(k)=\sqrt{k^{2}+M_{N}^{2}}+3\overline{\omega}^{0}$ (for
$\eta=0$) \cite{under2} where $M_{N}(\overline{\sigma})$ is the
effective nucleon mass in medium obtained by solving the bound state
Faddeev equation (\ref{fad}). The nucleon mass now is a complicated
function of scalar field $\overline{\sigma}$.  We define the baryon
density $\rho$ by
\begin{equation}
\rho=\gamma_{N}\int \frac{d^{3}k}{(2\pi)^3} n_{N}(k), 
\end{equation}
 where the Fermi distribution function $n_{N}(k)$ is at zero temperature the
 step function $n_{N}(k)=\Theta(k_{F}-k)$.

Imposing the self-consistence
 condition $\frac{\partial\Omega}{\partial\overline{\sigma}}=0$ leads
 to the following finite-density gap equation
\begin{equation}
\overline{\sigma}=2g_{\pi}\gamma_{q}i\int \frac{d^{4}k}{(2\pi)^4}\frac{M(k)f^{2}(k)}{k^{2}-M^{2}(k)}-g_{\pi}\gamma_{N}\int \frac{d^{3}k}{(2\pi)^3}\frac{M_{N}}{E_{N}(k)}\frac{\partial M_{N}}{\partial\overline{\sigma}} n_{N}(k). \label{sel1-n}
\end{equation}
This is a highly non-linear equation which is now coupled with the solutions of the Faddeev
equations. Therefore, in order to solve the gap equation at a given
density one needs to solve the Faddeev equation at the same time. In
this way, the non-perturbative feature of the nucleon as a composite
object is taken into account. This equation resembles the gap equation
in the vacuum Eq.~(\ref{sel1-0}). In order to obtain the
physical value of $\overline{\omega}_{0}$, we require that
$\frac{\partial\Omega}{\partial\overline{\omega}_{0}}=0$, for nuclear matter at rest this yields
\begin{equation}
\overline{\omega}_{0}=6g_{v}\gamma_{N}\int \frac{d^{3}k}{(2\pi)^{3}} n(k)=6g_{v}\rho.
\end{equation}

Notice that the general form of the Faddeev equation (\ref{fad}) for
the nucleon and the Bethe-Salpeter equation for the diquark in nuclear medium
remains the same as in the vacuum. The only new input is the in-medium
modified scalar mean field Eq.~(\ref{sel1-n}) which changes quark and
diquark propagators appearing in Eq.~(\ref{n9}). At low density we do
not have quark matter background, the Pauli-blocking is taken into
account in constructing the composite nucleon and in the nuclear matter \cite{under2,fad-new}.

\begin{figure}%[!tp]
%\vspace{.5cm}
      \includegraphics[height=.32\textheight] {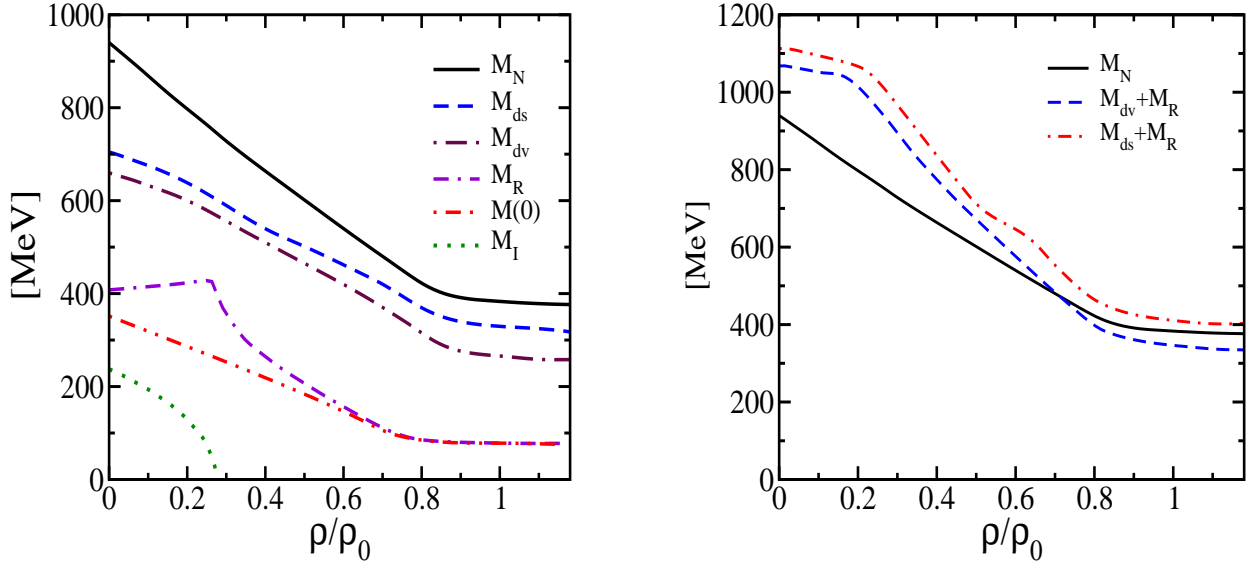}
       \caption{On the left the zero-momentum constituent quark mass
       $M(0)$ and the (real, imaginary) part of the first pole of the
       dressed quark propagator ($M_{R}$, $M_{I}$) are displayed as a
       function of density $\rho$ in the unit of the nuclear matter
       density $\rho_{0}$. We also show the effective nucleon $M_{N}$
       and the diquark masses $M_{ds}, M_{da}$ (for $g_{v}=0$) at
       finite density. On the right we show the fictitious scalar,  
       axial diquark-quark threshold $M_{ds}+M_{R}$ and $M_{dv}+M_{R}$, respectively
       as a function of density $\rho/\rho_{0}$ (the nucleon mass
       $M_{N}$ is also shown).}
\end{figure}

In Fig.~4 (left panel) we show the effective quark mass obtained as the solution of
the finite-density gap equation (\ref{sel1-n}), the scalar diquark mass
$M_{ds}$ and the axial-diquark mass $M_{da}$ at finite density. The
density dependent nucleon mass obtained from Faddeev Eq.~(\ref{fad})
is also shown. Notice that the nucleon mass does not follow from the
addition of the diquark and constituent quark masses because of the
relativistic Faddeev equation.

The effective constituent quark mass in the nuclear matter decreases
with a different slope compared to that in the quark matter
background \cite{non-s2}. This indicates that the many-body effect felt
by a constituent quark in the nucleon is different from the quark
matter. In Fig.~4 (left panel) we also show the real $M_{R}(\rho)$ and
imaginary $M_{I}(\rho)$ part of the first pole of the dressed quark
propagator as a function of nuclear matter density. The deconfinement
occurs where $M_{I}(\rho)=0$, this corresponds to
$\rho/\rho_{0}=0.28$. The real part of the first pole of the dressed
quark propagator $M_{R}$ increases with density up to the
deconfinement point and then it decreases and approaches the
zero-momentum constituent quark mass $M(0)$ near the nuclear matter density $\rho_{0}$. In our
model, we do not have a well-defined quark-diquark threshold. However,
one may define a fictitious quark-diquark threshold as $M_{ds}+M_{R}$
(or $M_{da}+M_{R}$), where $M_{R}$ is the real part of the first pole
of the dressed quark propagator.  It is noted from Fig~.4 (right
panel) that the nucleon mass remains below the fictitious scalar
diquark-quark threshold for all densities. However, it moves slightly
above the fictitious axial diquark-quark threshold near the
nuclear matter density $\rho_{0}$. Note that we show in Fig.~4 only the real
part of nucleon mass. Above the diquark-quark threshold nucleon pole
position is moved in the complex plane where the real part is taken as
physical mass. Our numerical procedure (see Appendix) is valid when
nucleon solution is not far above the fictitious threshold. It is natural to
assume that at some point as we increase the density, nucleons decay
to quarks and diquarks due to deconfinement. At the moment, we do not
know yet the order of a possible deconfinement transition and its
critical density. Based on our model approach, one can not address
such questions quantitatively.

The scalar and the axial-diquark masses in medium decrease in a very
similar way. Therefore, their relative importance in the nucleon
description remains almost the same at different densities. The
nucleon mass compared with the diquarks and the quarks has a bigger
decreasing slope at very low density. However, it tends to saturate
near the nuclear matter density. This effect is also observed for the
diquarks, however, less pronounced.

The stabilisation of the nucleon mass $M_{N}$ at about nuclear matter
density and the dynamical partial restoration of the chiral symmetry
may closely be interconnected. This can be realized, since after
chiral restoration of the constituent quark mass, the nucleon cannot
feel the presence of the nuclear medium any more. The medium is
incorporated via the quark propagator Eq.~(\ref{pro}) through the
finite-density gap equation (\ref{sel1-n}) into the diquark and
consequently in the diquark-quark Bethe-Salpeter equations. The
Pauli-blocking effect in the medium is not present for the quarks
since in the medium we do not have a quark matter background. We
employ the quarks only to construct the nucleons and assume that they
are not resolved by the medium. The stabilisation of the nucleon mass
in nuclear matter is the same as before.  In the local NJL model where
quark confinement is simulated by introducing an infrared cutoff the
stability of the nucleon mass is reached at higher density
\cite{under2}. In the context of the derivative scalar coupling model
this phenomenon is related to dynamical screening of the effective
coupling \cite{dsg}. In our model, it is derived from the non-locality of the underlying theory. 

\begin{figure}%[!tp]
      \includegraphics[height=.40\textheight]{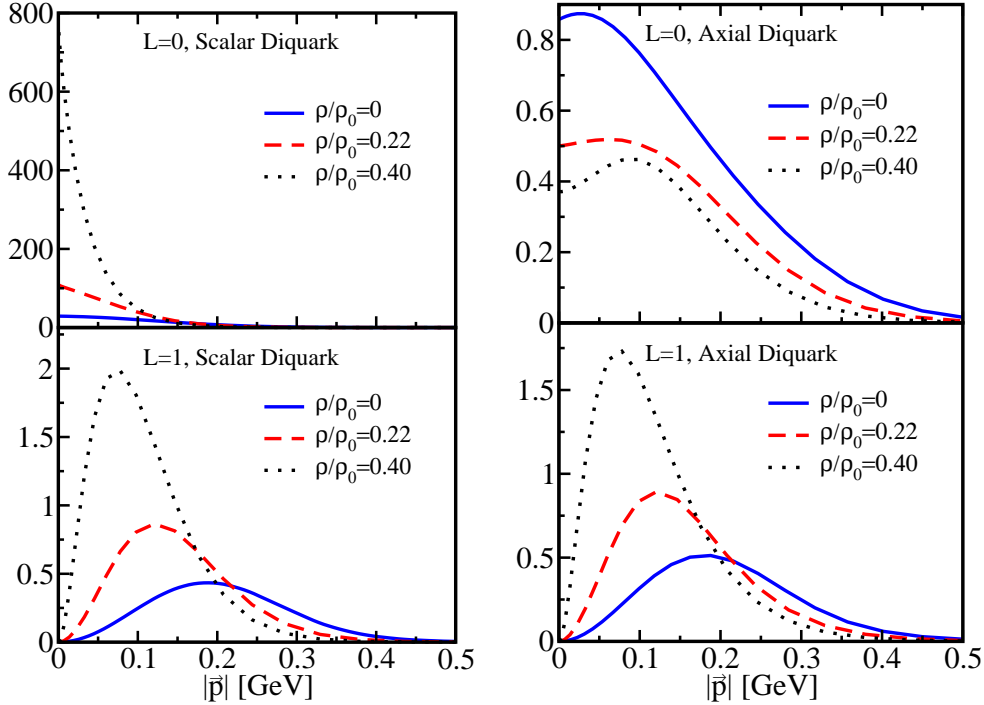}
      \caption{We show the density function of the scalar and axial
      diquark components within the nucleon at a given nuclear matter
      density $\rho$ with respect to relative momentum between diquark and
      quark. All given components have spin $S=1/2$ but different
      orbital angular momenta $L$ as indicated in the plot.}
\end{figure}

In order to find a better understanding of the internal structure of
the nucleon in medium, we construct the spectator quark momentum distribution function within
nucleon \cite{me1} as a function of nuclear matter density in terms of
various diquark components in the nucleon rest frame,
\begin{eqnarray}
\rho(|\vec{p}|,M_{N})&=&\int dp_{4}\psi^{\dag 5}(|\vec{p}|,p_{4},M_{N})(D^{5}(p_{d}))^{-1}\psi^{5}(|\vec{p}|,p_{4},M_{N})\nonumber\\
&+& \int
dp_{4}\psi^{\dag\mu\nu}(|\vec{p}|,p_{4},M_{N})(D^{\mu\alpha}(p_{d}))^{-1}\psi^{\alpha\nu}(|\vec{p}|,p_{4},M_{N}),\label{density-f}\
\end{eqnarray}
where $\psi^{\dag 5}$ and $\psi^{\mu\nu}$ are the Faddeev vertex
functions obtained from Eq.~(\ref{fad}) and $\vec{p}$ stands for the
space component of relative momentum $p$.  $D^{5}(p_{d}) $ and
$D^{\mu\nu}(p_{d})$ denote the diquark propagators.  Although this
definition is not unique, it provides some useful information about
the nature of compositeness of the nucleon and their response to the
medium. Note that the nucleon vertex function contains information
about the compositeness of the diquark through the diquark vertex function $\chi^{5,\mu}$, see
Eqs.~(\ref{new-ve},\ref{n9}).

In the rest frame the mass of the nucleon and its total angular
momentum are good quantum numbers. In the rest frame it is also
possible to decompose the nucleon vertex function in terms of tri-spinor
in different diquark channels each possessing definite orbital angular
momentum and spin which allows a direct interpretation of
different components \cite{o1,me1}. The results are shown in Fig.~5. It
is observed that at finite density the s-wave in the scalar diquark
channel is the dominant contribution to the nucleon ground state. As
the nuclear matter density increases the scalar and the axial diquark
masses decrease, consequently the diquark interaction couplings
$g_{s}$ and $g_{a}$ grow. This effect can be realized from Fig.~5,
since the maximum strength of the density function increases as we
increase the baryonic density. It is interesting to notice that from
Fig.~5 one can observe that the relative importance of the scalar
and axial diquarks remains intact in medium (at least at low
density). In order to estimate the change of the nucleon size in
response to nuclear matter environment, we compute
$p_{\text{RMS}}=(\langle \vec{p}^{2}\rangle-\langle
\vec{p}\rangle^{2})^{1/2}$ at finite density. This leads to
$p_{\text{RMS}} \text{(MeV)}=232, 182, 146$ for $\rho/\rho_{0}=0,
0.22$ and $0.40$, respectively. Therefore, as baryonic density
increases the nucleon size grows. An increase about $35\%$
is found for density $\rho/\rho_{0}\approx 0.40$.

Fig.~6 shows the binding energy per nucleon as a function of
density. The equation of state when the internal structure of nucleon
are taken into account in a presence of a vector field indeed
saturates. We adjust the vector mean field coupling $g_{v}$ in order
to have right binding energy per particle $E_{B}/A=-16$ MeV. This
corresponds to $\rho/\rho_{0}=0.60$ when $g_{v}=g_{\pi}/2.4$.  
%In Fig.~8, for comparison, we also show the equation of state without
%incorporating the internal structure of the nucleon for the same
%parameter set C and vector coupling $g_{v}=g_{\pi}/2.4$.  In this case
%the nuclear matter collapses. However, in this case it is still
%possible to obtain a saturating equation of state by increasing the
%vector-coupling, but the saturation point occurs at lower density with
%unrealistically too binding energy. A similar feature has been also
%reported in local NJL model when the quark structure of nucleon is
%neglected \cite{sa-1,sa-2}. 
%Therefore, taking into account the quark structure of the nucleon
%significantly improves the result. 
The fact that the equation of state curve cannot pass exactly through
the empirical saturation point implies that our hybrid approach based
on a mean field approximation in the medium is still a very crude
approximation\footnote{Notice that here we have only one free
parameter $g_{v}$ in medium and it is not straightforward to fix two
values via it.  A similar study by Bentz and Thomas in the context of
the local NJL model has shown that the binding energy saturates at too
high density \cite{under2}.}.

It is well known that in the chiral models the stability of
the nuclear matter depends very much on the dynamical chiral restoration
\cite{sa-1,sa-2,under2,lastamir}. 
The necessary condition for saturation of nuclear matter in any
relativistic mean field theory is that nucleon attraction mediated
from the $\sigma$-meson exchange should decrease at high density.  Therefore, if
$\sigma$-meson mass decreases too rapidly with density due to chiral
restoration, it may work against the stabilization of the system \cite{sa-1,sa-2,under2,lastamir}. However, as we have numerically proven,
it is in principle possible to obtain a saturating nuclear matter
equation of state, if the scalar field couples with the quarks instead
of directly with the nucleons (see also \cite{under2}). To this end, the nucleon is taken as a
diquark-quark state which moves in self consistent scalar and
vector fields coupling to the quarks.

\begin{figure}%[!tp]
      \includegraphics[height=.32\textheight]{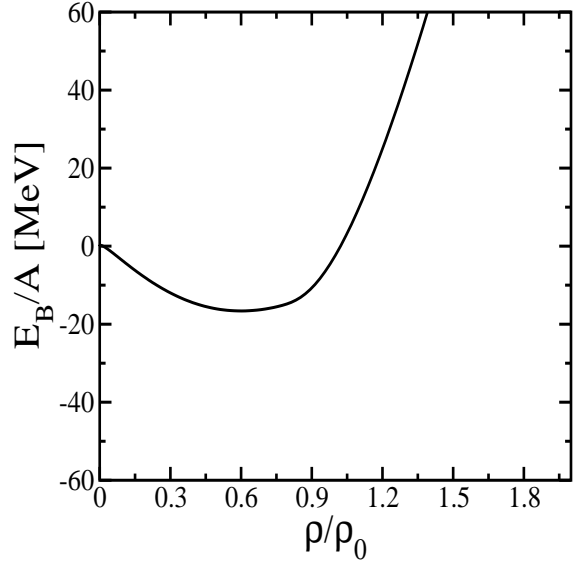}
      \caption{Energy per baryon as a function of $\rho/\rho_{0}$
      ($\rho_{0}$ is nuclear matter density), for parameter set C,
      when $g_{v}=g_{\pi}/2.4$.}
\end{figure}

\section{Conclusion}
In this paper we have investigated the nuclear matter in a non-local
NJL model.  We have worked out the equation of state of the nuclear
matter in a hybrid approximation.  First, we constructed the nucleon
as a bound state of a diquark and a quark in the relativistic Faddeev
approach and then built up nuclear matter by means of these composite
nucleons. In this way, we incorporated the internal quark structure of
the nucleon in a very simple framework.  We showed that in this
approximation the binding energy saturates. This is in contrast to the
local NJL model, where in a similar framework, a stable normal
nuclear matter does not exist unless one introduces a new extra
parameter into the model either by including an $8$-Fermi interaction
term \cite{sa-1} to induce the original coupling of the model density
dependent or an infrared cutoff to simulate the confinement effect
\cite{under2,under2-3}. Therefore, the long standing problem of matter
stability in the NJL model can be resolved by introducing non-locality
without invoking any adhoc new parameters. Note that the form factor
Eq.~(\ref{f-f}) does not introduce any new parameter except the cutoff
$\Lambda$ which is also present in the local NJL model.

We have also studied the nucleon properties such as the modification
of the nucleon mass and size in a nuclear matter medium. The nucleon
mass in the medium decreases very rapidly but saturates near the nuclear matter 
density $\rho_{0}$. We obtained the nucleon wave function from the Faddeev
equation and showed that the nucleon size significantly increases in
the medium.  This implies a swollen nucleon in nuclear matter which
has many attendant consequences \cite{bag2,bag1,last1,last2}. Our
estimation of in-medium nucleon size should be taken more
qualitatively since the quark density function given in
Eq.~(\ref{density-f}) has not a unique definition. We found that the
swelling of the nucleon is about $35\%$ at about half nucleon matter
density. One should note that in our model at nuclear matter
saturation density $\rho=0.6\rho_{0}$ the ``confinement'' mechanism is
no longer at work, see Fig.~4 (left panel) and the nucleon becomes close to the
fictitious diquark-quark threshold, see Fig.~4 (right panel).  Moreover, since the nucleon mass
saturates at high density, the nucleon will not expand
forever\footnote{We could not obtain the Faddeev wave function at the
higher density due to computation difficulties to verify this
effect.}. Other studies have predicted different values\footnote{One
should be aware that nucleon size has not generally a unique
definition among various models.}. For an example, it has been shown
within the Friedberg-Lee nontopological-soliton bag model that the
swelling of the nucleon is about $30\%$ at normal nuclear matter
density
\cite{bag2}. In the Skyrmion picture nucleon swelling about $40\%$ has
been reported \cite{last1}. It has been also shown that in the
quark-meson coupling model nucleon swelling about $25-35\%$ in
saturated nuclear matter can explain the observed depletion of the
structure function in the medium Bjorken $x$ region \cite{last2}.

We have also investigated the role of diquarks within the nucleon in the
nuclear matter medium. The scalar and axial diquark masses decrease with
a very similar slope in the medium. Despite the fact that the nucleon mass
decreases (and its size increases) significantly, the role of the scalar
and the axial diquark in the nucleon description remains almost the same in the nuclear medium.

In this paper, for simplicity we assumed that the auxially vector
field to be local. The role of the vector meson in medium is still an
open question and it deserves further investigation in our model as well. 

\section*{Acknowledgements}
AHR acknowledges the financial support from the Alexander von Humboldt
foundation. AHR would like to thank M. Beyer and D. Blaschke for useful discussions.

\appendix*
\section{Numerical method}
There are some subtleties involved in evaluating loop integrals as
of Eq.~(\ref{lo1}) in Euclidean space. For simplicity this integral is evaluated at the
timelike momentum $q=(0,iq_{0})$ in Euclidean space. For the confining parameter set,
each quark propagator has a pair of complex conjugate poles. As we
increase $q_{0}$, these poles in $S(p_{\pm})$ (where $p_{\pm}=p\pm \frac{1}{2}q$) have a chance to cross the real axis and
may produce an imaginary part in the meson (or diquark) propagator when
a threshold for decay of a meson (or diquark) into $qq$- unphysical
states opens\footnote{The quark propagator has many set of quartets of
complex poles and for the deconfinement case we have purely real poles
in form of doublets. The external momentum going into the $qq$-loops
(or $q\bar{q}$-loops) does not change the structure of the quark
propagator poles, however it changes their positions. In other words,
if the relation Eq.~(\ref{n4}) is satisfied with the parameters of the
model, poles will always remain in form of quartets \cite{pb}.}, see Fig.~9.  Therefore, the standard Wick rotation of the
integration contour cannot be applied. We use the prescription
proposed by Cutkosky {\em et al.} \cite{cutk} and successfully applied to such
models in Refs.~\cite{pb,a1,u}. This amounts to a deformation of the
integration contour (as indicated in Fig.~10) to ensure that the meson or diquark
propagator does not develop an imaginary part, when the first pair of
complex poles of quark propagator cross the real axis. At the pinch point, both the naive integral over Euclidean
four-momentum and the residue contribution diverge, although these
divergences cancel to leave a finite result \cite{pb,cutk}. 
\begin{figure}%[!tp]
      \includegraphics[height=.20\textheight]{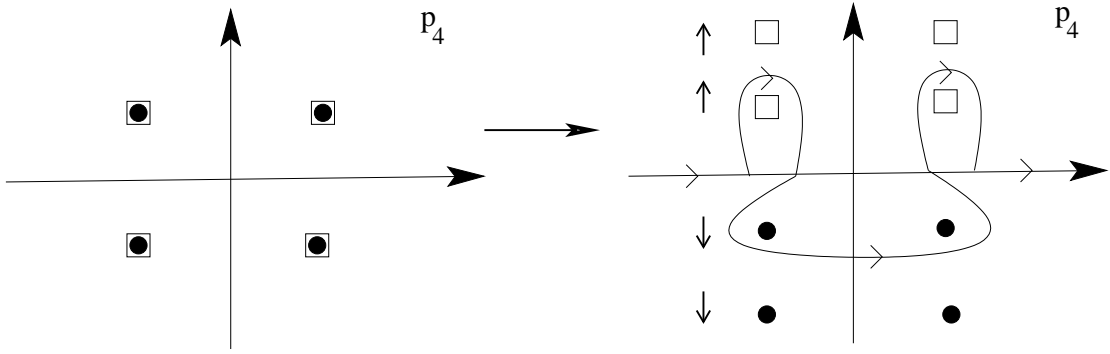} \caption{We
      show the first complex conjugate poles of loop integrals in the
      $p_{4}$ plane. On the left panel, the pole positions at
      $q_{0}=0$ are shown. The open boxes and filled circles denote
      the poles of the $p_{-}$ and $p_{+}$, respectively. On the right
      panel, we show the deformed integration contour beyond the pinch
      point. }
\end{figure}

The diquark-quark loops in the Faddeev equation (\ref{fad}) are
computed in the rest frame of the nucleon $P=(E,0)$. First, we
identify the singularities involved in the kernel of Faddeev equation
(\ref{n9}) with respect to the integration momentum $k_{0}$.  The
singularities from quark propagator will occur at :
\begin{eqnarray}
&&\text{pole 1}: \hspace{1cm} k_{0}=-\eta E+\omega^{q}_{k}-i\epsilon, \\
&&\text{pole 2}: \hspace{1cm} k_{0}=-\eta E-\omega^{q}_{k}+i\epsilon,\
\end{eqnarray}
where $\omega^{q}_{k}=\sqrt{k^{2}+M_{R}^{2}}$. The singularities from diquark propagator occurs:
\begin{eqnarray}
&&\text{pole 3}: \hspace{1cm} k_{0}=(1-\eta)E+\omega^{d}_{k}-i\epsilon, \\
&&\text{pole 4}: \hspace{1cm} k_{0}=(1-\eta)E-\omega^{d}_{k}+i\epsilon,\
\end{eqnarray}
where we define $\omega^{d}_{k}=\sqrt{k^{2}+M_{d}^{2}}$ and $M_{d}$ is the
the lower-energy pole in the diquark $T$-matrix (the mass of the
lighter diquark). The exchanged quark propagator is also singular at:
\begin{eqnarray}
&&\text{pole 5}: \hspace{1cm} k_{0}=-p_{0}+\omega^{q}_{p+k}+(1-2\eta)E-i\epsilon, \\
&&\text{pole 6}: \hspace{1cm} k_{0}=-p_{0}-\omega^{q}_{p+k}+(1-2\eta)E+i\epsilon,\
\end{eqnarray}
with $\omega^{q}_{p+k}=\sqrt{(k+p)^{2}+M_{R}^{2}}$. Note that all masses in the above equations are
density dependent. We define 
\begin{equation}
\alpha(\rho)=(1-\eta)M_{R}(\rho)-\eta M_{d}(\rho) \hspace{1cm} \eta\in [0,1].
\end{equation}
It is easy to verify that in the complex $k_{0}$, the poles $2$ and
$4$ lie always left of $k_{0}=\alpha(\rho)$ and poles $1$ and $3$ lie
always right of $k_{0}=\alpha$ for $E<M_{R}+M_{d}$.  To avoid the
singularities of the kernel, we perform a Wick rotation at a given
density on the energy variables around $k_{0}=\alpha(\rho)$,
i. e. $k_{0}\to\alpha(\rho)+ik_{1}$
($-\infty<k_{1}<+\infty$). Although $\alpha$ changes with density, it
is in principle independent of the Mandelstam parameter $\eta$. This
provides an internal check on numerical procedure. In
Refs.~\cite{njln3,walet}, the parameter $\eta$ is taken $1/2$ from the
outset. The location of the quark exchanged poles $5$ and $6$ depends
on the $p_{0}$. One needs to bend the integration contour in such a
way that the pole $5$ and $6$ to lie right and left of the path,
respectively. To this end, we continue the momentum $p_{0}$ to complex
plane with $p_{0}=\alpha(\rho)+ik_{2}$ then the poles $5$ and $6$ lie
always right and left of the integration path
$k_{0}=\alpha(\rho)+ik_{1}$.  Above the diquark-quark threshold the
solution of the effective Faddeev equation should be searched in
complex $P^{2}$-plane. In this case the complex pole coordinates
$P^{2}=E^{2}-i\gamma$ are interpreted as a mass, $E$, and $\gamma$ is
related to a corresponding width \cite{com-m}. Our numerical method is valid when
the location of the complex pole is close to the real axis and we
concentrate only on computation of the mass. In practice, one may
increase the stability of the numerical procedure by taking advantage
of $\eta$ (in order to escape the above-mentioned poles far enough
away). This implies that one can find a plateau within the range
$0<\eta<1$ where the results do not depend on the parameter $\eta$
\cite{me1,o1,o2}. The resulting complex coupled integral equations are
solved by using the algorithm introduced by Oettel {\em et al.}
\cite{o1,o2} (see also Ref.\cite{me1}).


\begin{thebibliography}{20}
\bibitem{njl}
For a review see:  S. P. Klevansky, Rev. Mod. Phys. {\bf 64}, 649
(1992). 
\bibitem{sa-1}
V. Koch, T. S. Biro, J. Kunz and U. Mosel, Phys. Lett. {\bf B185}, 1 (1987). 
\bibitem{sa-1-1}
J. Da Providencia, M. C. Ruivo and C. A. De Sousa,
Phys. Rev. {\bf D36}, 1882 (1987). 
\bibitem{sa-2}
M. Buballa, Nucl. Phys. {\bf A611}, 393 (1996).
\bibitem{under2}
 W. Bentz and A. W. Thomas, Nucl. Phys. {\bf A696}, 138 (2001). 
\bibitem{sa-3}
 B. D. Serot and J. D.  Walecka, Int. J. Mod. Phys. {\bf E6},515
 (1997); Adv. Nucl. Phys. {\bf 16},1 (1986).
\bibitem{delta1}
For an example: T. Schafer, E. V. Shuryak, J. J. M. Verbaarschot, Nucl. Phys. {\bf B412},143 (1994).
\bibitem{njln3}
 N. Ishii, W. Bentz, K. Yazaki, Nucl. Phys. {\bf A587},617 (1995).
\bibitem{p1}
 O. Nachtmann and H. J. Pirner, Z. Phys. {\bf C21}, 277 (1984). 

\bibitem{sa-4}
J. R. Smith, G. A. Miller, Phys. Rev. {\bf C65}, 055206 (2002);  Phys. Rev. Lett. {\bf 91}, 212301 (2003).
\bibitem{fad-new}
I.C. Cloet, W. Bentz and A. W. Thomas, Phys. Rev. Lett. {\bf 95}, 052302 (2005).
 \bibitem{non1}
 D. Diakonov, V. Yu. Petrov, Sov. Phys. JETP {\bf 62}, 204 (1985); Nucl. Phys. {\bf B245}, 259 (1984);
 D. Diakonov, V. Yu. Petrov, P. V. Pobylitsa, Nucl. Phys. {\bf B306}, 809 (1988);
I. V. Anikin, A. E. Dorokhov and L. Tomio, Phys. Part. Nucl. {\bf 31}, 509 (2000).
\bibitem{pb}
R. D. Bowler and M. C. Birse, Nucl. Phys. {\bf A582}, 655 (1995);
R. S. Plant and M. C. Birse, Nucl. Phys. {\bf A628}, 607 (1998).
\bibitem{lat-m}
 J. Skullerud, D. B. Leinweber and A. G. Williams, Phys. Rev. {\bf D64}, 074508 (2001).
\bibitem{pb-2}
R. S. Plant and M. C. Birse, Nucl. Phys. {\bf A703}, 717 (2002).
\bibitem{a1}
 A. Scarpettini, D. G. Dumm and N. N. Scoccola, Phys. Rev. {\bf D69}, 114018 (2004). 
\bibitem{a3}
 W. Broniowski, B. Golli and  G. Ripka, Nucl. Phys. {\bf A703}, 667 (2002);
 B. Golli, W. Broniowski and G. Ripka, Phys. Lett. {\bf B437}, 24 (1998).  
 \bibitem{me1}
 A. H. Rezaeian, N. R. Walet and M. C. Birse, Phys. Rev. {\bf C70}, 065203 (2004).
\bibitem{non-s2}
 D. G. Dumm and N. N. Scoccola, Phys. Rev. {\bf D65}, 074021 (2002).
\bibitem{a2}
I. General, D. G. Dumm and N. N. Scoccola, Phys. Lett. {\bf B506},
267 (2001); D. G. Dumm and N. N. Scoccola, Phys. Rev. {\bf C72}, 014909 (2005);
 R. S. Duhau, A. G. Grunfeld and N. N. Scoccola,  Phys. Rev. {\bf D70}, 074026 (2004). 

\bibitem{njln1}
A. Buck, R. Alkofer and H. Reinhardt, Phys. Lett. {\bf B286}, 29 (1992);
S. Huang and J. Tjon, Phys. Rev. {\bf C49}, 1702 (1994); H. Asami, N. Ishii, W. Bentz and K. Yazaki, Phys. Rev. {\bf C51}, 3388 (1995);
C. Hanhart and S. Krewald, Phys. Lett. {\bf B344}, 55 (1995).

\bibitem{o1}
M. Oettel, G. Hellstern, R. Alkofer and H. Reinhardt, Phys. Rev. {\bf C58}, 2459, (1998).
\bibitem{o2} 
G. Hellstern, R. Alkofer, M. Oettel and H. Reinhardt, Nucl. Phys. {\bf A627}, 679 (1997);
M. Oettel, R. Alkofer and L. von Smekal, Eur. Phys. J. {\bf A8}, 553 (2000);
S. Ahlig, R. Alkofer, C.S. Fischer, M. Oettel, H. Reinhardt and H. Weigel, Phys. Rev. {\bf D64}, 014004 (2001);
M. Oettel, L. Von Smekal, R. Alkofer, Comput. Phys. Commun. {\bf 144}, 63 (2002).
\bibitem{me2}
A. H. Rezaeain, hep-ph/0507304. 
\bibitem{asy}
 C. D. Roberts and A. G. Williams, Prog. Part. Nucl. Phys. {\bf 33}, 477 (1994);
 C. D. Roberts and S. M. Schmidt, Prog. Part. Nucl. Phys. {\bf 45}, S1 (2000).
\bibitem{non-bir}
 H. Ito, W. W. Buck and F. Gross, Phys. Lett. {\bf B248}, 28 (1990);
 H. Ito, W. Buck and F. Gross, Phys. Rev. {\bf C43}, 2483 (1991).
\bibitem{sep-i}
T. Sch\"afer and E. V. Shuryak, Rev. Mod. Phys. {\bf 70}, 323 (1998);
D. Diakonov, Prog. Part. Nucl. Phys. {\bf 36}, 1 (1996). 
\bibitem{gluon}
G. V. Efimov and S. N. Nedelko, Phys. Rev. {\bf D51}, 176 (1995); Y. V. Burdanov, G. V. Efimov, S. N. Nedelko and S. A. Solunin, Phys. Rev. {\bf D54}, 4483 (1996).

\bibitem{sde-p}
D. Atkinson and D. W. E. Blatt, Nucl. Phys. {\bf B151}, 342 (1979).
\bibitem{sde-new}
P. Maris and H. A. Holties, Int. J. Mod. Phys. {\bf A7}, 5369 (1992);
S. J. Stainsby and R. T. Cahill, Int. J. Mod. Phys. {\bf A7}, 7541 (1992);
P. Maris, Phys. Rev. {\bf D50}, 4189 (1994).
\bibitem{sde-qe}
 P. Maris, Phys. Rev. {\bf D52}, 6087 (1995).
\bibitem{qcdsum}
 H.G. Dosch and S. Narison, Phys. Lett.{\bf B417},173 (1998).
\bibitem{lat-con}
 L. Giusti, F. Rapuano, M. Talevi and A. Vladikas, Nucl. Phys. {\bf B538}, 249 (1999).
\bibitem{bub1}
M. Buballa,  Phys. Rept. {\bf 407},205 (2005). 
\bibitem{bub2}
 M. Lutz, S. Klimt and  W. Weise, Nucl. Phys. {\bf A542},  521 (1992).  
\bibitem{pir}
J. Meyer, G. Papp, H. J. Pirner, T. Kunihiro, Phys. Rev. {\bf C61}, 035202 (2000).
\bibitem{dsg}
J. Zimanyi and S. A. Moszkowski, Phys. Rev. {\bf C42}, 1416 (1990);
 R. Aguirre and A. L. De Paoli, Eur. Phys. J. {\bf A13}, 501 (2002).
\bibitem{bag1}
For examples: J. V. Noble, Phys. Rev. Lett. {\bf 46}, 412 (1981);
F. E. Close, R. G. Roberts, G. G. Ross, Phys. Lett. {\bf B129}, 346 (1983);
M. Jandel,G. Peters, Phys. Rev. {\bf D30}, 1117 (1984);
R. L. Jaffe, F. E. Close, R. G. Roberts, G. G. Ross,  Phys. Lett. {\bf B134}, 449 (1984);
L. S. Celenza, A. Rosenthal and C. M. Shakin, Phys. Rev. {\bf C31}, 232 (1985);
G. Chanfray, H. J. Pirner, Phys. Rev. {\bf C35}, 760 (1987). 
\bibitem{bag2}
M. Jandel, G. Peters, Phys. Rev. {\bf D30}, 1117 (1984).
\bibitem{last1}
M. Rho,  Phys. Rev. Lett. {\bf 54}, 767 (1985). 
\bibitem{last2}
X. Jin, B. K. Jennings,  Phys. Rev. {\bf C54}, 1427 (1996); Phys. Rev.{\bf C55}, 1567 (1997).
\bibitem{cutk}
R. E. Cutkosky, P. V. Landshoff, D. I. Olive and J. C. Polkinghorne, Nucl. Phys. {\bf B12}, 281 (1969).
\bibitem{u}
M. S. Bhagwat, M. A. Pichowsky and P.C. Tandy, Phys. Rev. {\bf D67}, 054019 (2003).
\bibitem{walet}
S. Pepin, M. C. Birse, J. A. McGovern and N. R. Walet, Phys. Rev. {\bf C61}, 055209 (2000).
\bibitem{lastamir}
A. H. Rezaeian, nucl-th/0512027.
\bibitem{under2-3}
 I. C. Cloet, W. Bentz and A. W. Thomas, Phys. Rev. Lett. {\bf 95}, 052302 (2005).
\bibitem{com-m}
 V. Bernard, A. H. Blin, B. Hiller, Y. P. Ivanov, A. A. Osipov and U. G. Meissner, Phys. Lett. {\bf B409}, 483 (1997).
\end{thebibliography}
\end{document}